\documentclass[%
 reprint,
 amsmath,amssymb,
 aps,
]{revtex4-2}

\usepackage{graphicx}
\usepackage{dcolumn}
\usepackage{bm}
\usepackage{physics}

\begin{document}

\preprint{APS/123-QED}

\title{Density Matrices of Pulsed Quantum Light}

\author{Joscelyn van der Veen}
  \email{joscelyn.vanderveen@utoronto.ca}
\author{Daniel F. V. James}%
\affiliation{%
 Department of Physics, University of Toronto, 60 St. George St., Toronto, ON M5S1A7, Canada}%

\date{\today}

\begin{abstract}
Pulsed light is represented by density matrices whose correlation functions have different properties of stationarity and coherence. We show that these density matrices yield equivalent measurement results after accounting for finite detection times. This in turn reveals a physical meaning to the length of a quantization cavity and explains the origin of the physical spectrum for pulsed light. By generalizing the density matrices for pulsed Gaussian states of light, we can describe pulsed light of varying statistics in the same basis and relate the photocount rate to the pulse spectrum.
\end{abstract}

\maketitle

\textit{Introduction---}Pulsed light has a vast array of applications ranging from medicine and machining to fundamental physics \cite{strickland2018}. In quantum optics, the high peak powers of pulsed light make it advantageous for driving nonlinear processes, such as spontaneous parametric downconversion (SPDC) and optical parametric amplification (OPA) \cite{boyd2008}. These nonlinear processes are essential for quantum communications \cite{Osorio2012} and metrology \cite{Aasi2013} to achieve their promised potential. 

Despite the ubiquity of pulsed light sources, namely mode-locked lasers, their theoretical description remains incomplete. Measurement techniques make the assumption that light has a ``coherence'' property \cite{Alonso2024} even though the master equation (ME) used to describe mode-locked lasers does not account for coherent effects \cite{Perego2020}. Moreover, the assumption of an often undefined ``coherence'' property complicates the development of theory to describe pulsed light with different statistics. This is not only a foundational problem for optics but also  impedes development of promising novel techniques such as intensity interferometery with pulsed thermal light \cite{Horoshko2025} and squeezing in high harmonic generation \cite{Gorlach2020, Theidel2024}.

The difficulty with generalizing theory beyond coherence assumptions is rooted in understanding the statistics of light. Despite coherence theory being established in the 1960s \cite{Glauber1963-quantumcoherence,Glauber1963,Mandel1965} there was a great debate in the 2000s on the coherence of a single mode (SM) laser \cite{Molmer1997,GeaBanacloche1998,Rudolph2001,wiseman2001}. This debate questioned whether a SM laser should be described by a coherent state, viz:

\begin{equation}
    \hat{\rho}_{coh}\equiv\ket{\alpha}\bra{\alpha}=e^{-|\alpha|^2}\sum_{n,m=0}^\infty\frac{\alpha^n(\alpha^*)^m}{\sqrt{n!m!}}\ket{n}\bra{m},
    \label{eq:rho-sm-coh}
\end{equation}

\noindent where $\alpha\in\mathbb{C}$ is an eigenvalue of the annihilation operator, or a Poisson-distributed phase-random mixture of number states, 

\begin{equation}
    \hat{\rho}_{num}=e^{-|\alpha|^2}\sum_{n=0}^\infty\frac{|\alpha|^{2n}}{n!}\ket{n}\bra{n}.
    \label{eq:rho-sm-num}
\end{equation}

It was eventually shown that both density matrices can represent a SM laser by considering the phase of an optical mode as an extrinsic property \cite{Bartlett2006}. 

However, the debate only considered SM fields where the correlation function, from which the properties of light are typically defined \cite{gerry2004}, for both Eq. (\ref{eq:rho-sm-coh}) and Eq. (\ref{eq:rho-sm-num}) are the same. This is no longer the case when we extend the question of the density matrix to pulsed light, which is multimode. 

Currently, the theoretical descriptions of pulsed quantum light do not agree on a particular model density matrix. For example, in some works pulsed light is represented by a pure state multifrequency density matrix, usually transformed into the basis of the fundamental frequency distribution \cite{Christ2011,Theidel2024,Lemieux2025}. In other works the behaviour of the pulse is considered at a single frequency, implicitly assuming a density matrix where each frequency independently conforms to a specific statistical model \cite{Manceau2019,Stammer2022}. Other authors assume that time dependence can be incorporated into the initial density matrix of the field in vacuum \cite{Gorlach2020,Lamprou2025}. 

In this letter, we show that when we account for the finite measurement time of photodetectors, the different model density matrices are indistinguishable. In fact, they are limiting cases of a generalized density matrix with indistinguishable photocount rate that depends on a phase randomness parameter whose value we can choose to best simplify future calculations. We show that the finite measurement time of photodetectors means that the $n$th order photocount rate for pulsed light is independent of this parameter, thus making the density matrices indistinguishable in measurements. We extend the generalization of the multimode density matrix to other Gaussian states like squeezed vacuum.

\textit{Poissonian Light---}Light sources like lasers obey Poissonian statistics \cite{Scully1967,MandelandWolf} so we begin by considering a Poissonian density matrix. In a single mode, the most general form of a Poissonian density matrix is,

\begin{equation}
    \hat{\rho}_{p,sm}(\lambda)=e^{-\lambda}\sum_{n,m=0}^\infty\frac{(\sqrt{\lambda})^{n+m}}{\sqrt{n!m!}}q_{nm}\ket{n}\bra{m},
    \label{eq:rho-sm-poisson}
\end{equation}

\noindent where $\lambda$ is the Poisson rate parameter and $q_{nn}=1$.

The $n$th order correlation function is defined in general as \cite{Glauber2006},

\begin{multline}
    G^{(n)}(t_1,\ldots,t_n,t_n',\ldots,t_1') \\
    =\Tr{\hat{\rho}\hat{E}^{(-)}(t_1)\ldots\hat{E}^{(-)}(t_n)\hat{E}^{(+)}(t_{n}')\ldots\hat{E}^{(+)}(t_1')},
    \label{eq:Gn}
\end{multline}

\noindent where $\hat{E}^{(+)}(t)$ is the electric field operator projected along some polarization direction $\boldsymbol{\mu}$ at an arbitrary position (we can define as $\boldsymbol{r}=0$). For clarity, we will consider correlation functions for beam-like propagating light which is described by the one-dimensional electric field operator, 

\begin{equation}
    \hat{E}^{(+)}(t)=i\mathcal{V}\sum_k\sqrt{\frac{k}{L}}\hat{a}_ke^{-ikct}.
    \label{eq:E+}
\end{equation}

\noindent where $\mathcal{V}=\sqrt{\hbar c/2\epsilon_0 A}$ for beam cross-section $A$ is on the order of $10^{-5}$V for a 1mm beam radius.

We obtain the electric field operator by assuming that free space acts as a 1D cavity of length $L$. This imposes periodic boundary conditions used to quantize the field into discrete modes with wavenumber $k=2\pi m /L$ for $m\in\mathbb{Z}$. Normally, we would take $L\to\infty$ at the end of the calculation because measurements do not depend on the distance a beam propagates in free space \cite{gerry2004}. However, we will keep $L$ explicit because we will find that it has a physical meaning for pulsed light. 

With the single mode Poissonian density matrix (Eq. (\ref{eq:rho-sm-poisson})), the correlation functions all reduce to a simple form,

\begin{multline}
    G^{(n)}(t_1,\ldots,t_n,t_n',\ldots,t_1') \\
    = \left(\frac{\mathcal{V}^2k}{L}\right)^ne^{-ick(t_1'+\ldots+t_n'-t_1-\ldots-t_n)}\lambda^{n}.
\end{multline}

We now wish to consider the density matrix for pulsed light. Since a single frequency mode field is always harmonic we must introduce more modes for the bandwidth of pulsed light. Light from a mode-locked laser can be filtered with a bandpass filter of arbitrary bandwidth and retain Poissonian statistics once it is above threshold \cite{Roche2023}. This implies that each frequency mode is independently Poissonian distributed. Thus, we have a factorizable multimode Poissonian density matrix,

\begin{equation}
    \hat{\rho}_p=\bigotimes_k\hat{\rho}_{k,p}(\lambda_k),
    \label{eq:rho-p}
\end{equation}

\noindent where the density matrix for each mode $\hat{\rho}_{k,p}(\lambda_k)$ is given by Eq. (\ref{eq:rho-sm-poisson}).

To find the first order correlation function, we substitute Eq. (\ref{eq:rho-p}) into Eq. (\ref{eq:Gn}) obtaining a double sum over $\Tr{\hat{\rho}_p\hat{a}_k^\dagger\hat{a}_{k'}}$. We can separate the correlation function into two sums: a double sum over the trace when the modes are not equal ($k\neq k'$) and a single sum over the trace when the modes are equal ($k=k'$). By adding and subtracting the value of the first term when $k=k'$, we can arrive at the first order correlation function,

\begin{multline}
    G^{(1)}(t,t)=\mathcal{V}^2\sum_k\left(\frac{k}{L}\right)\left(1-|\zeta_k|^2\right)\lambda_k \\
    +\mathcal{V}^2\left|\sum_k\sqrt{\frac{k}{L}}e^{-ickt}\zeta_k\sqrt{\lambda_k}\right|^2
    \label{eq:G1-p-1},
\end{multline}

\noindent where we define 

\begin{equation}
    \zeta_k\equiv e^{-\lambda_k}\sum_n\frac{\lambda_k^n}{n!}q^{(k)}_{n+1,n},
    \label{eq:zeta}
\end{equation}

\noindent which is a convergent series since $|q^{(k)}_{nm}|\leq 1$ for all modes $k$.

Recall that $\lambda_k$ is the Poisson rate parameter that has the physical interpretation of the expected number of photons measured in some mode.  The $\zeta_k$ are not associated with a physical quantity and are introduced by the construction of the density matrix. They thus act similarly to a choice of gauge: the $\zeta_k$ can be chosen to simplify calculations. If we let the magnitude of $\zeta_k$ be constant and equal across all modes, i.e. $\zeta_k\equiv\xi$ and we combine the Poisson rate parameter and the phase of $\zeta_k$ into

\begin{equation}
    \alpha_k\equiv\sqrt{\lambda_k}\exp\left(i\arg\{\zeta_k\}\right),
    \label{eq:alpha_k}
\end{equation}

\noindent then we obtain the correlation function in terms of the usual Fourier coefficients $\alpha_k$.

With these definitions, the first order correlation function has the form,

\begin{multline}
    G^{(1)}(t,t)=\left(1-\xi^2\right)\sum_k\mathcal{V}^2\left(\frac{k}{L}\right)|\alpha_k|^2 \\
    +\xi^2\left|\mathcal{V}\sum_k\sqrt{\frac{k}{L}}e^{-ickt}\alpha_k\right|^2.
    \label{eq:G1-p}
\end{multline}

The choice of $\xi$ can yield either Eq. (\ref{eq:rho-sm-coh}) ($\xi=1$) or Eq. (\ref{eq:rho-sm-num}) ($\xi=0$). We can interpret the choice of $\xi$ as an assumption of phase reference. 

The correlation function Eq. (\ref{eq:G1-p}) varies in two of the primary properties used to distinguish light: stationarity and coherence. Light is considered stationary (in the wide sense) if the first order correlation function is independent of the origin of time \cite{MandelandWolf,Goodman_2015}, like in the first term of Eq. (\ref{eq:G1-p}). Most classical statistical optics results are derived for stationary light \cite{MandelandWolf,Goodman_2015} and these must be extended in varying ways to describe pulsed light \cite{Koivurova2024,Schoonover2009}. Light is considered coherent if the correlation function is separable in the time arguments \cite{Glauber1963}, like in the second term of Eq. (\ref{eq:G1-p}). Based on these definitions, different values of $\xi$ should describe light that behaves very differently. To determine whether this is the case, we must consider the quantity that is actually measured: the photocount rate of light. 

If the time between two measurements can be arbitrarily small, then the photocount rate is proportional to the correlation function. However, the time between two measurements at a single ideal photodetector is fundamentally limited by the finite response time of a medium. The time between measurements at two different photodetectors also cannot be arbitrarily close because whenever two photodetectors are discriminated in time, we must consider higher order correlations. 

When considering the finite response time, the photocount rate is proportional to the time average of the correlation function rather than the correlation function \cite{Glauber2006,Milonni1994,MandelandWolf}. The photocount rate is thus,

\begin{equation}
    w^{(1)}(T)=\frac{\eta}{T}\int_0^T G^{(1)}(t,t)\dd{t},
\end{equation}

\noindent where $\eta$ is the detector's quantum efficiency \cite{Glauber2006}, which we assume to be constant for all relevant wavelengths. 

Whenever the light is stationary, which is always the case for both single mode and thermal light, then $G^{(1)}(t,t)$ is constant and the photocount rate is proportional to the first order correlation function with $t_1=t_2$.  

\begin{figure}
    \centering
    \includegraphics[width=0.9\linewidth]{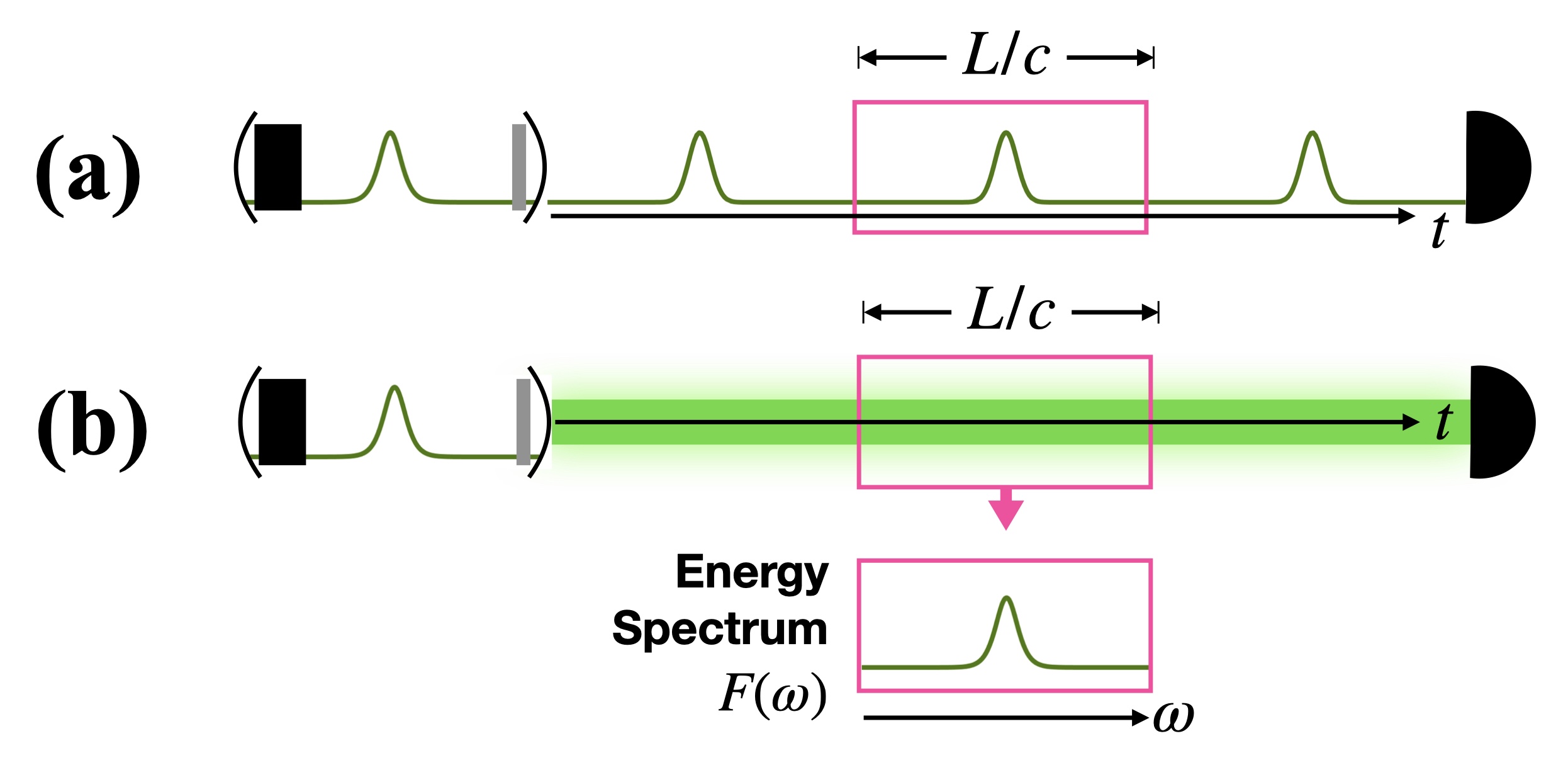}
    \caption{We quantize the electric field in some cavity of length $L$. A coherent density matrix (a) describes the field as a pulse train where each pulse is separated by a time $L/c$. A stationary density matrix (b) instead has the field existing at all times but with a spectrum that is the Fourier spectrum of a single pulse in any time period $L/c$.}
    \label{fig:quant-len}
\end{figure}

Consider now the photocount rate for the Poissonian density matrix, which is the time average of Eq. (\ref{eq:G1-p}). The photocount rate has a stationary incoherent part (first term) and a nonstationary coherent part (second term). By recalling that $k=2\pi m/L$ for $m\in\mathbb{Z}$, we can write the nonstationary coherent part (i.e. the second term on the RHS of Eq. (\ref{eq:G1-p})) in terms of the Fourier series,

\begin{equation}
    \mathcal{E}(t)=\mathcal{V}\sum_m\sqrt{\frac{2\pi m}{L^2}}e^{-2\pi imt/(L/c)}\alpha_{2\pi m/L},
    \label{eq:f-series}
\end{equation}

\noindent which is repeated in time with a period of $L/c$. Thus the absolute value $|\mathcal{E}(t)|$ gives the pulse train amplitude (see Fig. \ref{fig:quant-len}(a)). 

If we define the number of periods (pulses) in the interval $[0,T]$ as $N$ and the shape of a single pulse as $f(t)$, i.e. $\mathcal{E}(t)=\sum_{n=-\infty}^\infty f(t-nT)$, then the coherent part of the first order photocount rate for the Poissonian density matrix depends on,

\begin{equation}
    \int_0^T|\mathcal{E}(t)|^2\dd{t}=N\int_{-\infty}^\infty|f(t)|^2\dd{t}.
\end{equation}

Unlike the nonstationary coherent part of the photocount rate, the stationary incoherent part ((i.e. the first term on the RHS of Eq. (\ref{eq:G1-p})) is time independent. This means we view the field as being present at all times instead of coming in pulses (see Fig. \ref{fig:quant-len}(b)). We describe the stationary incoherent part instead in terms of a spectrum per quantum of energy $\hbar \omega$, where $\omega$ is a continuous frequency. To write the stationary incoherent part in terms of a continuous frequency distribution, consider the inverse of the Fourier series $\mathcal{E}(t)$ from the nonstationary coherent part,

\begin{equation}
    \mathcal{V}\sqrt{\frac{k}{L}}\alpha_k=\frac{c}{L}\int_{-L/2c}^{L/2c}\mathcal{E}(t)e^{ickt}\dd{t}.
    \label{eq:inv-f-series}
\end{equation}

We can substitute Eq. (\ref{eq:inv-f-series}) into the stationary incoherent part of the photocount rate and use the fact that a sum of complex exponentials is a Dirac comb to write it as,

\begin{equation}
    \mathcal{V}^2\sum_k\left(\frac{k}{L}\right)|\alpha_k|^2 \\
    =\frac{c}{L}\int_{-L/2c}^{L/2c}|\mathcal{E}(t)|^2\dd{t}.
\end{equation}

The magnitude squared of the Fourier series $\mathcal{E}(t)$ over a single time period is just the amplitude of a single pulse, whose shape we have already defined as $f(t)$. Since $f(t)$ is the shape of a single pulse, it has a Fourier transform that is a continuous frequency distribution over $\omega$, which we can call $F(\omega)$. Thus, the stationary spectrum per quanta of energy for a pulsed field is the Fourier spectrum of a single pulse. We can find this same result by taking the sum to an integral while requiring normalization of the cavity.

The ratio of the measurement time $T$ to $L/c$ is the important quantity of this calculation. Typically, when quantizing the electric field in free space to arrive at the operator $\hat{E}(t)$ (Eq. (\ref{eq:E+})), we take the side length $L$ to be arbitrarily large. However, light propagating in free space still must be generated from a source. In the case of a pulsed laser, that source is a cavity. We can consider the frequencies of a single laser pulse to be quantized with respect to the laser cavity length (we refer here to the optical path length of the cavity which is twice the distance between the two reflecting surfaces in a cavity). The pulses are emitted in a train separated in time by the cavity round-trip time $\tau_{rep}$ \cite{milonni2010}. This equates to a spatial separation of the pulses in vacuum. We can therefore visualize the stationary incoherent part of the photocount rate as in Fig. \ref{fig:quant-len}(b): the field exists at all time but has a frequency distribution given by the pulse spectrum in any time period $L/c$ such that a measurement time $T$ includes $N$ copies of $L/c$. Thus $T/(L/c)$ is the number of pulses $N$ that contribute to the measurements. 

We thus find that any multimode independently-Poissonian-distributed density matrix (Eq. (\ref{eq:rho-p})) where all $\alpha_k$ correspond to the Fourier coefficients (i.e. Eq. (\ref{eq:alpha_k}) holds) can describe the photocount rate of pulsed light as,

\begin{equation}
    w^{(1)}=\frac{\eta}{\tau_{rep}}\int_{-\infty}^\infty|f(t)|^2\dd t=\frac{\eta}{2\pi \tau_{rep}}\int_{-\infty}^\infty|F(\omega)|^2\dd{\omega},
    \label{eq:I1-p}
\end{equation}

\noindent where 

\begin{equation}
    \frac{1}{\tau_{rep}}=\frac{N}{T}=\frac{c}{L}.
\end{equation}

\textit{Higher order photocount rate---}To determine whether multimode independently-Poissonian-distributed density matrices are truly indistinguishable regardless of stationarity and coherence, we must determine the higher order photocount rate measured by some $n$ photodetectors. The photocount rate at multiple photodetectors is dependent on higher order correlation functions. For $n$ detectors, the $n$-th order photocount rate is the counting rate per (unit time)$^n$ \cite{Glauber2006}.

To retain $\alpha_k$ as Fourier coefficients whose magnitude is the average photon number and which contain the complex phase associated with the field, we must specialize the general form of the Poissonian density matrix to,

\begin{equation}
    \hat{\rho}_{k,p}(\alpha_k)=e^{-|\alpha_k|^2}\sum_{n,m=0}^\infty\frac{\alpha_k^n(\alpha_k^*)^m}{\sqrt{n!m!}}\xi^{|n-m|}\ket{n}\bra{m},
    \label{eq:rho-spec}
\end{equation}

\noindent where $\xi$ is the same phase randomness factor from the first order correlation function, which yields a coherent state when $\xi=1$ and a stationary mixture of number states when $\xi\to0$.

We find that in general for any density matrix of the form Eq. (\ref{eq:rho-spec}), the $n$-th order photocount rate is independent of $\xi$ (see App. \ref{app:poiss-I-2} for the explicit calculation of the second order photocount rate),

\begin{equation}
    w^{(n)}=\left[\frac{\eta}{\tau_{rep}}\int_{-\infty}^\infty|f(t)|^2\dd{t}\right]^n.
    \label{eq:I-n-poiss}
\end{equation}

Thus, we find the principal result of this letter: \textit{Poissonian-distributed density matrices are equivalent in measurement, regardless of their stationarity or coherence, meaning we can choose the most convenient model density matrix}. 

\textit{Other Gaussian States---}We can generalize the equivalence of density matrices to other Gaussian states. For example, we can write the density matrices for squeezed vacuum as,

\begin{multline}
    \hat{\rho}_s=\bigotimes_k\frac{1}{\cosh(r_k)}\sum_{n_k,m_k=0}^\infty (-\tanh(r_k))^{n_k+m_k} \\
    \times e^{i\phi_k(n_k-m_k)}\frac{\sqrt{(2n_k)!(2m_k)!}}{2^{n_k+m_k}n_k!m_k!}\xi^{|n_k-m_k|}\ket{2n_k}\bra{2m_k}.
    \label{eq:rho-s}
\end{multline}

This density matrix will also yield the same $n$-th order photocount rate for all $\xi$ (see App. \ref{app:squeeze-I-2}). 

As with Poissonian light, we can identify the stationary spectrum of pulsed squeezed light as the Fourier spectrum of a single pulse in the time period $L/c$ such that the first order photocount rate can also be given by Eq. (\ref{eq:I1-p}). This relationship between the stationary spectrum and a single pulse spectrum allows us to relate expected higher order intensities with the easily measureable quantity of the spectrum. For example, the second order photocount rate for squeezed vacuum is (see App. \ref{app:squeeze-I-2}),

\begin{multline}
    w^{(2)}=\left(\frac{\eta}{2\pi\tau_{rep}}\right)^2\left[\left(\int_{-\infty}^\infty|F(\omega)|^2\dd{\omega}\right)^2\right. \\
    + \frac{2}{2\pi\tau_{rep}}\int_{-\infty}^\infty|F(\omega)|^4\dd{\omega} \\
    +\left.\left(\frac{2\pi\mathcal{V}^2}{c^2}\right)\int_{-\infty}^\infty\omega|F(\omega)|^2\dd{\omega}\right].
    \label{eq:I-2-squeeze}
\end{multline}

This differs from the photocount rate for Poissonian light and can thus be used to distinguish squeezed light without requiring photon number resolution.

\textit{Conclusion---}We have shown the equivalence of density matrices whose properties of stationarity and coherence differ. Through the process of showing equivalence, we have arrived at the physical meaning of the quantization cavity length: the cavity length of the physical process that generates traveling light. We can thus describe not only thermal light but also pulsed Poissonian and squeezed light by a multimode density matrix that is a product of mixtures of number states in different modes. The calculation of higher order intensities in terms of the pulse spectrum also provides a method for distinguishing the statistics of pulsed light from straightforward spectrum and photocount rate measurements.

\textit{Acknowledgements---}We acknowledge the support of the Natural Sciences and Engineering Research Council of Canada (NSERC) and Vice Dean, Research \& Infrastructure of the University of Toronto.

\bibliography{apssamp}

@article{Bartlett2006,
author = {Bartlett, Stephen D. and Rudolph, Terry and Spekkens, Rober W.},
title = {DIALOGUE CONCERNING TWO VIEWS ON QUANTUM COHERENCE: FACTIST AND FICTIONIST},
journal = {International Journal of Quantum Information},
volume = {04},
number = {01},
pages = {17-43},
year = {2006},
doi = {10.1142/S0219749906001591},
URL = {https://doi.org/10.1142/S0219749906001591},
eprint = {https://doi.org/10.1142/S0219749906001591}
}

@book{MandelandWolf,
  title = {Optical Coherence and Quantum Optics},
  ISBN = {9781139644105},
  url = {http://dx.doi.org/10.1017/CBO9781139644105},
  DOI = {10.1017/cbo9781139644105},
  publisher = {Cambridge University Press},
  author = {Mandel,  Leonard and Wolf,  Emil},
  year = {1995},
  month = sep 
}

@article{Molmer1997,
  title = {Optical coherence: A convenient fiction},
  author = {M\o{}lmer, Klaus},
  journal = {Phys. Rev. A},
  volume = {55},
  issue = {4},
  pages = {3195--3203},
  numpages = {0},
  year = {1997},
  month = {Apr},
  publisher = {American Physical Society},
  doi = {10.1103/PhysRevA.55.3195},
  url = {https://link.aps.org/doi/10.1103/PhysRevA.55.3195}
}

@inbook{Glauber2006,
  publisher = {John Wiley \& Sons, Ltd},
  author = {Glauber, Roy J.},
  isbn = {9783527610075},
  title = {Optical Coherence and Photon Statistics},
  booktitle = {Quantum Theory of Optical Coherence},
  chapter = {2},
  pages = {23-182},
  doi = {https://doi.org/10.1002/9783527610075.ch2},
  url = {https://onlinelibrary.wiley.com/doi/abs/10.1002/9783527610075.ch2},
  eprint = {https://onlinelibrary.wiley.com/doi/pdf/10.1002/9783527610075.ch2},
  year = {2006}
}

@article{Mandel1965,
  title = {Coherence Properties of Optical Fields},
  author = {Mandel, L. and Wolf, E.},
  journal = {Rev. Mod. Phys.},
  volume = {37},
  issue = {2},
  pages = {231--287},
  numpages = {0},
  year = {1965},
  month = {Apr},
  publisher = {American Physical Society},
  doi = {10.1103/RevModPhys.37.231},
  url = {https://link.aps.org/doi/10.1103/RevModPhys.37.231}
}

@article{GeaBanacloche1998,
  title = {Comment on “Optical coherence: A convenient fiction”},
  volume = {58},
  ISSN = {1094-1622},
  url = {http://dx.doi.org/10.1103/PhysRevA.58.4244},
  DOI = {10.1103/physreva.58.4244},
  number = {5},
  journal = {Physical Review A},
  publisher = {American Physical Society (APS)},
  author = {Gea-Banacloche,  Julio},
  year = {1998},
  month = Nov,
  pages = {4244–4246}
}

@article{Rudolph2001,
  title = {Requirement of Optical Coherence for Continuous-Variable Quantum Teleportation},
  volume = {87},
  ISSN = {1079-7114},
  url = {http://dx.doi.org/10.1103/PhysRevLett.87.077903},
  DOI = {10.1103/physrevlett.87.077903},
  number = {7},
  journal = {Physical Review Letters},
  publisher = {American Physical Society (APS)},
  author = {Rudolph,  Terry and Sanders,  Barry C.},
  year = {2001},
  month = {July}
}

@misc{wiseman2001,
      title={Comment on ``Requirement of optical coherence for continuous--variable quantum teleportation'' by Terry Rudolph and Barry C. Sanders}, 
      author={H. M. Wiseman},
      year={2001},
      eprint={quant-ph/0104004},
      archivePrefix={arXiv},
      primaryClass={quant-ph},
      url={https://arxiv.org/abs/quant-ph/0104004}, 
}

@book{Goodman_2015, 
  place={Hoboken, NJ}, 
  title={Statistical Optics}, 
  publisher={John Wiley and Sons Inc},
  author={Goodman, Joseph W.}, 
  year={2015}
}

@book{Milonni1994,
  title = {The Quantum Vacuum: An Introduction to Quantum Electrodynamics},
  author = {Milonni, Peter W.},
  ISBN = {9780080571492},
  url = {http://dx.doi.org/10.1016/C2009-0-21295-5},
  DOI = {10.1016/c2009-0-21295-5},
  publisher = {Elsevier},
  year = {1994}
}

@book{boyd2008,
  title={Nonlinear Optics},
  author={Boyd, R.W. and Prato, D.},
  isbn={9780080485966},
  lccn={2008271820},
  url={https://books.google.ca/books?id=uoRUi1Yb7ooC},
  year={2008},
  publisher={Academic Press}
}

@article{Lamprou2025,
  title = {Nonlinear Optics Using Intense Optical Coherent State Superpositions},
  volume = {134},
  ISSN = {1079-7114},
  url = {http://dx.doi.org/10.1103/PhysRevLett.134.013601},
  DOI = {10.1103/physrevlett.134.013601},
  number = {1},
  journal = {Physical Review Letters},
  publisher = {American Physical Society (APS)},
  author = {Lamprou,  Th. and Rivera-Dean,  J. and Stammer,  P. and Lewenstein,  M. and Tzallas,  P.},
  year = {2025},
  month = Jan 
}

@article{Gorlach2020,
  title = {The quantum-optical nature of high harmonic generation},
  volume = {11},
  ISSN = {2041-1723},
  url = {http://dx.doi.org/10.1038/s41467-020-18218-w},
  DOI = {10.1038/s41467-020-18218-w},
  number = {1},
  journal = {Nature Communications},
  publisher = {Springer Science and Business Media LLC},
  author = {Gorlach,  Alexey and Neufeld,  Ofer and Rivera,  Nicholas and Cohen,  Oren and Kaminer,  Ido},
  year = {2020},
  month = Sept 
}

@article{Christ2011,
  title = {Probing multimode squeezing with correlation functions},
  volume = {13},
  ISSN = {1367-2630},
  url = {http://dx.doi.org/10.1088/1367-2630/13/3/033027},
  DOI = {10.1088/1367-2630/13/3/033027},
  number = {3},
  journal = {New Journal of Physics},
  publisher = {IOP Publishing},
  author = {Christ,  Andreas and Laiho,  Kaisa and Eckstein,  Andreas and Cassemiro,  Katiúscia N and Silberhorn,  Christine},
  year = {2011},
  month = Mar,
  pages = {033027}
}

@article{Lemieux2025,
  title = {Photon bunching in high-harmonic emission controlled by quantum light},
  volume = {19},
  ISSN = {1749-4893},
  url = {http://dx.doi.org/10.1038/s41566-025-01673-6},
  DOI = {10.1038/s41566-025-01673-6},
  number = {7},
  journal = {Nature Photonics},
  publisher = {Springer Science and Business Media LLC},
  author = {Lemieux,  Samuel and Jalil,  Sohail A. and Purschke,  David N. and Boroumand,  Neda and Hammond,  T. J. and Villeneuve,  David and Naumov,  Andrei and Brabec,  Thomas and Vampa,  Giulio},
  year = {2025},
  month = May,
  pages = {767–771}
}

@article{Theidel2024,
  title = {Evidence of the Quantum Optical Nature of High-Harmonic Generation},
  volume = {5},
  ISSN = {2691-3399},
  url = {http://dx.doi.org/10.1103/PRXQuantum.5.040319},
  DOI = {10.1103/prxquantum.5.040319},
  number = {4},
  journal = {PRX Quantum},
  publisher = {American Physical Society (APS)},
  author = {Theidel,  David and Cotte,  Viviane and Sondenheimer,  René and Shiriaeva,  Viktoriia and Froidevaux,  Marie and Severin,  Vladislav and Merdji-Larue,  Adam and Mosel,  Philip and Fr\"{o}hlich,  Sven and Weber,  Kim-Alessandro and Morgner,  Uwe and Kovacev,  Milutin and Biegert,  Jens and Merdji,  Hamed},
  year = {2024},
  month = Nov 
}

@article{Manceau2019,
  title = {Indefinite-Mean Pareto Photon Distribution from Amplified Quantum Noise},
  volume = {123},
  ISSN = {1079-7114},
  url = {http://dx.doi.org/10.1103/PhysRevLett.123.123606},
  DOI = {10.1103/physrevlett.123.123606},
  number = {12},
  journal = {Physical Review Letters},
  publisher = {American Physical Society (APS)},
  author = {Manceau,  Mathieu and Spasibko,  Kirill Yu. and Leuchs,  Gerd and Filip,  Radim and Chekhova,  Maria V.},
  year = {2019},
  month = Sept 
}

@article{Stammer2022,
  title = {Theory of entanglement and measurement in high-order harmonic generation},
  volume = {106},
  ISSN = {2469-9934},
  url = {http://dx.doi.org/10.1103/PhysRevA.106.L050402},
  DOI = {10.1103/physreva.106.l050402},
  number = {5},
  journal = {Physical Review A},
  publisher = {American Physical Society (APS)},
  author = {Stammer,  Philipp},
  year = {2022},
  month = Nov 
}

@book{milonni2010,
  title={Laser Physics},
  author={Milonni, Peter W. and Eberly, Joseph H.},
  isbn={9780470387719},
  url={https://onlinelibrary.wiley.com/doi/book/10.1002/9780470409718},
  year={2010},
  publisher={John Wiley \& Sons}
}

@book{gerry2004,
  title={Introductory Quantum Optics},
  author={Gerry, Christopher and Knight, Peter},
  year={2004},
  publisher={Cambridge University Press},
  address={Cambridge, UK},
  isbn={978-0521820356}
}

@article{Glauber1963,
  title = {Coherent and Incoherent States of the Radiation Field},
  volume = {131},
  ISSN = {0031-899X},
  url = {http://dx.doi.org/10.1103/PhysRev.131.2766},
  DOI = {10.1103/physrev.131.2766},
  number = {6},
  journal = {Physical Review},
  publisher = {American Physical Society (APS)},
  author = {Glauber,  Roy J.},
  year = {1963},
  month = Sept,
  pages = {2766–2788}
}

@article{Osorio2012,
  title = {Heralded photon amplification for quantum communication},
  author = {Osorio, C. I. and Bruno, N. and Sangouard, N. and Zbinden, H. and Gisin, N. and Thew, R. T.},
  journal = {Phys. Rev. A},
  volume = {86},
  issue = {2},
  pages = {023815},
  numpages = {4},
  year = {2012},
  month = {Aug},
  publisher = {American Physical Society},
  doi = {10.1103/PhysRevA.86.023815},
  url = {https://link.aps.org/doi/10.1103/PhysRevA.86.023815}
}

@article{Alonso2024,
  title = {Space–time characterization of ultrashort laser pulses: A perspective},
  volume = {9},
  ISSN = {2378-0967},
  url = {http://dx.doi.org/10.1063/5.0219447},
  DOI = {10.1063/5.0219447},
  number = {7},
  journal = {APL Photonics},
  publisher = {AIP Publishing},
  author = {Alonso,  Benjamín and D\"{o}pp,  Andreas and Jolly,  Spencer W.},
  year = {2024},
  month = July 
}

@article{Perego2020,
  title = {Coherent master equation for laser modelocking},
  volume = {11},
  ISSN = {2041-1723},
  url = {http://dx.doi.org/10.1038/s41467-019-14013-4},
  DOI = {10.1038/s41467-019-14013-4},
  number = {1},
  journal = {Nature Communications},
  publisher = {Springer Science and Business Media LLC},
  author = {Perego,  Auro M. and Garbin,  Bruno and Gustave,  Fran\c{c}ois and Barland,  Stephane and Prati,  Franco and de Valcárcel,  Germán J.},
  year = {2020},
  month = Jan 
}

@article{Aasi2013,
    	author = {Aasi, J. and others},
	journal = {Nature Photonics},
	number = {8},
	pages = {613--619},
	title = {Enhanced sensitivity of the LIGO gravitational wave detector by using squeezed states of light},
	volume = {7},
	year = {2013}
}

@article{Horoshko2025,
  title = {Time-resolved second-order autocorrelation function of parametric down-conversion},
  author = {Horoshko, Dmitri B. and Srivastava, Shivang and So\ifmmode \acute{s}\else \'{s}\fi{}nicki, Filip and Miko\l{}ajczyk, Micha\l{} and Karpi\ifmmode \acute{n}\else \'{n}\fi{}ski, Micha\l{} and Brecht, Benjamin and Kolobov, Mikhail I.},
  journal = {Phys. Rev. A},
  volume = {112},
  issue = {2},
  pages = {023703},
  numpages = {13},
  year = {2025},
  month = {Aug},
  publisher = {American Physical Society},
  doi = {10.1103/7ckm-tm3r},
  url = {https://link.aps.org/doi/10.1103/7ckm-tm3r}
}

@article{Glauber1963-quantumcoherence,
  title = {The Quantum Theory of Optical Coherence},
  author = {Glauber, Roy J.},
  journal = {Phys. Rev.},
  volume = {130},
  issue = {6},
  pages = {2529--2539},
  numpages = {0},
  year = {1963},
  month = {Jun},
  publisher = {American Physical Society},
  doi = {10.1103/PhysRev.130.2529},
  url = {https://link.aps.org/doi/10.1103/PhysRev.130.2529}
}

@misc{strickland2018,
  author       = {Strickland, Donna},
  title        = {Generating High-Intensity Ultrashort Optical Pulses},
  howpublished = {Nobel Lecture, Nobel Prize Outreach 2026},
  year         = {2018},
  url          = {https://www.nobelprize.org/prizes/physics/2018/strickland/lecture/},
  note         = {Accessed: 2026-08-07}
}

@article{Scully1967,
  title = {Quantum Theory of an Optical Maser. I. General Theory},
  volume = {159},
  ISSN = {0031-899X},
  url = {http://dx.doi.org/10.1103/PhysRev.159.208},
  DOI = {10.1103/physrev.159.208},
  number = {2},
  journal = {Physical Review},
  publisher = {American Physical Society (APS)},
  author = {Scully,  Marlan O. and Lamb,  Willis E.},
  year = {1967},
  month = July,
  pages = {208–226}
}

@article{Koivurova2024,
  title = {Nonstationary optics: tutorial},
  volume = {41},
  ISSN = {1520-8532},
  url = {http://dx.doi.org/10.1364/JOSAA.516951},
  DOI = {10.1364/josaa.516951},
  number = {4},
  journal = {Journal of the Optical Society of America A},
  publisher = {Optica Publishing Group},
  author = {Koivurova,  Matias and Laatikainen,  Jyrki and Friberg,  Ari T.},
  year = {2024},
  month = Mar,
  pages = {615}
}

@article{Schoonover2009,
  title = {The generalized Wolf shift for cyclostationary fields},
  volume = {17},
  ISSN = {1094-4087},
  url = {http://dx.doi.org/10.1364/OE.17.004705},
  DOI = {10.1364/oe.17.004705},
  number = {6},
  journal = {Optics Express},
  publisher = {Optica Publishing Group},
  author = {Schoonover,  Robert W. and Davis,  Brynmor J. and Carney,  P. Scott},
  year = {2009},
  month = Mar,
  pages = {4705}
}

@article{Roche2023,
  title = {Decoherence and Turbulence Sources in a Long Laser},
  volume = {131},
  ISSN = {1079-7114},
  url = {http://dx.doi.org/10.1103/PhysRevLett.131.053801},
  DOI = {10.1103/physrevlett.131.053801},
  number = {5},
  journal = {Physical Review Letters},
  publisher = {American Physical Society (APS)},
  author = {Roche,  Amy and Slepneva,  Svetlana and Kovalev,  Anton and Pimenov,  Alexander and Vladimirov,  Andrei G. and Giudici,  Massimo and Marconi,  Mathias and Huyet,  Guillaume},
  year = {2023},
  month = Aug 
}

\newpage

\appendix

\section{Second order photocount rate with a Poissonian density matrix} \label{app:poiss-I-2}

The second order photocount rate is the time integrated second order correlation function,

\begin{multline}
    w^{(2)}=\frac{\eta^2}{T^2}\mathcal{V}^4\sum_{k_1k_2}\sum_{k_1'k_2'}\int_0^T\dd{t_1}\int_0^T\dd{t_2}\frac{\sqrt{k_1k_2k_1'k_2'}}{L^2} \\
    \times e^{ic(k_1-k_1')t_1}e^{ic(k_2-k_2')t_2}\Tr{\hat{\rho}\hat{a}_{k_1}^\dagger\hat{a}_{k_2}^\dagger\hat{a}_{k_1'}\hat{a}_{k_2'}}
    \label{eq:I-2}
\end{multline}

To evaluate the photocount rate, we separate the correlation function into sums with different combinations of equal modes. For the second order correlation function, there are 15 combinations of the mode indices $k_1,k_2,k_1',$ and $k_2'$:

\begin{align}
    \begin{split}
        &k_1\neq k_2\neq k_1'\neq k_2' \\
        &k_1= k_2\neq k_1'\neq k_2'  \\
        &\ldots \\
        &k_1=k_2=k_1'=k_2'
    \end{split}
\end{align}

When we take the density matrix $\hat{\rho}$ to be the multimode Poissonian density matrix $\hat{\rho}_p$ (Eq. (\ref{eq:rho-spec})), the expectation value of normally ordered field operators is given by,

\begin{equation}
    \Tr{\hat{\rho}_{k,p}(\alpha_k)(\hat{a}_k^\dagger)^r(\hat{a}_k)^s}=(\alpha_k^*)^r\alpha_k^s\xi^{|r-s|}.
    \label{eq:p-trace}
\end{equation}

We can use Eq. (\ref{eq:p-trace}) to evaluate the traces over normally ordered field operators with equal modes. We then add and subtract sums with equal modes from those where the modes are not equal such that each sum is over all modes, i.e. with $\sum_{k_2'\neq k_1,k_2}$ we add and subtract the terms with $k_2'=k_1$, $k_2'=k_2$, and $k_2'=k_1=k_2$. When we integrate over time in the photocount rate we get an expression that involves the Fourier series $\mathcal{E}(t)$ (Eq. (\ref{eq:f-series})) and its inverse (Eq. (\ref{eq:inv-f-series})). If we write the photocount rate in terms of $\mathcal{E}(t)$, we can group it into eight terms,

\begin{widetext}
    \begin{multline}
        w^{(2)}\frac{T^2}{\eta^2}=\xi^4\left(\int_0^T|\mathcal{E}(t)|^2\dd{t}\right)^2+2(\xi^2-\xi^4)\frac{cT}{L}\int_{-L/2c}^{L/2c}|\mathcal{E}(t)|^2\dd{t}\int_0^T|\mathcal{E}(t)|^2\dd{t} \\
        + 2(\xi^2-\xi^4)\sum_k\mathcal{V}^2\left(\frac{k}{L}\right)|\alpha_k|^2\left|\int_0^T\mathcal{E}(t)e^{ickt}\dd{t}\right|^2 -2(\xi^2-\xi^4)\sum_k\mathcal{V}^2\left(\frac{k}{L}\right)|\alpha_k|^2\left(\mathcal{V}\sqrt{\frac{k}{L}}\alpha_k^*\int_0^T\mathcal{E}(t)e^{ickt}\dd{t}+c.c.\right) \\
        +2(\xi^2-\xi^4)\sum_k\mathcal{V}^2\left(\frac{k}{L}\right)|\alpha_k|^2\left|\mathcal{V}\sqrt{\frac{k}{L}}\alpha_k\right|^2+(1-\xi^4-2(\xi^2-\xi^4))\left(\frac{cT}{L}\int_{-L/2c}^{L/2c}|\mathcal{E}(t)|^2\dd{t}\right)^2 \\
    +(1-\xi^4-2(\xi^2-\xi^4))\int_0^T\int_0^T\left|\frac{c}{L}\int_{-L/2c}^{L/2c}\int_{-L/2c}^{L/2c}\mathcal{E}(t)\mathcal{E}^*(t')\sum_me^{2\pi i m(t-t'+t_1-t_2)/(L/c)}\dd{t}\dd{t'}\right|^2 \dd{t_1}\dd{t_2} \\
    -(1-\xi^4-2(\xi^2-\xi^4))\left(\frac{cT}{L}\right)^2\int_{-\frac{L}{2c}}^{\frac{L}{2c}}\int_{-\frac{L}{2c}}^{\frac{L}{2c}}\int_{-\frac{L}{2c}}^{\frac{L}{2c}}\int_{-\frac{L}{2c}}^{\frac{L}{2c}}\mathcal{E}(t)\mathcal{E}^*(t')\mathcal{E}^*(t'')\mathcal{E}(t''')\sum_me^{2\pi im(t+t'''-t'-t'')/(L/c)}\dd{t}\dd{t'}\dd{t''}\dd{t'''}
    \end{multline}
\end{widetext}

The sum of the third, fourth, and fifth terms is zero. The sum of the seventh and eighth terms is also zero once we write the sum of complex exponentials as a Dirac comb and integrate. The first, second, and sixth terms can all be written in terms of the shape of the single pulse $f(t)$ and the number of pulses $N$. We thus find,

\begin{equation}
    w^{(2)}\frac{T^2}{\eta^2}=N^2\left(\int_{-\infty}^\infty|f(t)|^2\right)^2
\end{equation}

\section{Photocount rate of squeezed vacuum}\label{app:squeeze-I-2}

To calculate the first and second order intensities with the general squeezed vacuum density matrix Eq. (\ref{eq:rho-s}), we can find the following traces of normally ordered field operators \cite{gerry2004},

\begin{align}
    \Tr{\hat{\rho}_s\hat{a}_k}&=0 \\
    \Tr{\hat{\rho}_s\hat{a}_k\hat{a}_k}&=-\xi e^{i\phi_k}\sinh(r_k)\cosh(r_k) \\
    \Tr{\hat{\rho}_s\hat{a}_k^\dagger\hat{a}_k}&=\sinh^2(r_k) \\
    \Tr{\hat{\rho}_s\hat{a}_k^\dagger\hat{a}_k\hat{a}_k} &= 0 \\
    \Tr{\hat{\rho}_s\hat{a}_k^\dagger\hat{a}_k^\dagger\hat{a}_k\hat{a}_k} &= \cosh^2(r_k)\sinh^2(r_k)+2\sinh^4(r_k)
\end{align}

The first order correlation function is stationary and clearly independent of $\xi$ so the first order photocount rate is simply,

\begin{equation}
    w^{(1)}_s=\eta \sum_k\mathcal{V}^2\left(\frac{k}{L}\right)\sinh^2(r_k)
\end{equation}

We still wish to describe the stationary spectrum in terms of a spectrum per quantum of energy. By comparing the squeezed vacuum photocount rate to the photocount rate for a multimode Poissonian density matrix, we can see that the summand is given by the inverse Fourier series, 

\begin{equation}
    \mathcal{V}\sqrt{\frac{k}{L}}\sinh(r_k)=\frac{c}{L}\int_{-L/2c}^{L/2c}\mathcal{E}_s(t)e^{ickt}\dd{t}.
    \label{eq:inv-f-series-squeeze}
\end{equation}

\noindent where $|\mathcal{E}_s(t)|^2$ is a train of pulses with amplitude $|f(t)|^2$ such that the first order photocount rate is,

\begin{equation}
    w^{(1)}=\frac{\eta}{\tau_{rep}}\int_{-\infty}^\infty|f(t)|^2\dd t=\frac{\eta}{2\pi \tau_{rep}}\int_{-\infty}^\infty|F(\omega)|^2\dd{\omega}
\end{equation}

This allows us to write higher order intensities in terms of single pulse spectra. For the second order photocount rate, we have,

\begin{widetext}
\begin{multline}
    w^{(2)}=\eta^2\left(\mathcal{V}^2\sum_k\left(\frac{k}{L}\right)\sinh^2(r_k)\right)^2 + \eta^2(1-\xi^2)\sum_k\mathcal{V}^4\left(\frac{k}{L}\right)^2\sinh^2(r_k)\cosh^2(r_k)\\
    +\frac{\eta^2}{T^2}\int_0^T\dd{t_1}\int_0^T\dd{t_2}\left[\left|\sum_k\mathcal{V}^2\left(\frac{k}{L}\right)\sinh^2(r_k)e^{-ick(t_2-t_1)}\right|^2
    +\xi^2\left|\sum_k\mathcal{V}^2\left(\frac{k}{L}\right)e^{i\phi_k}
    \sinh(r_k)\cosh(r_k)e^{-ick(t_1+t_2)}\right|^2\right]
\end{multline}
\end{widetext}

Comparing to the calculation of the second order photocount rate for the Poissonian density matrix, we note that in general, 

\begin{equation}
    \int_0^T\dd{t_1}\int_0^T\dd{t_2}\left|\sum_kA_ke^{-ick(t_2\pm t_1)}\right|^2=T^2\sum_k|A_k|^2
\end{equation}

We thus find the second order photocount rate,

\begin{multline}
    w^{(2)}=\eta^2\left(\sum_k\mathcal{V}^2\left(\frac{k}{L}\right)\sinh^2(r_k)\right)^2 \\
    + 2\eta^2\sum_k\mathcal{V}^4\left(\frac{k}{L}\right)^2\sinh^4(r_k) \\
    +\eta^2\sum_k\mathcal{V}^4\left(\frac{k}{L}\right)^2\sinh^2(r_k)
\end{multline}

\noindent which can be given in terms of the spectrum of a single pulse $F(\omega)$ as,

\begin{multline}
    w^{(2)}=\left(\frac{\eta}{2\pi\tau_{rep}}\right)^2\left[\left(\int_{-\infty}^\infty|F(\omega)|^2\dd{\omega}\right)^2\right. \\
    + \left.\frac{2}{2\pi\tau_{rep}}\int_{-\infty}^\infty|F(\omega)|^4\dd{\omega} +\left(\frac{2\pi\mathcal{V}^2}{c^2}\right)\int_{-\infty}^\infty\omega|F(\omega)|^2\dd{\omega}\right]
\end{multline}

\end{document}